\documentclass[sigconf]{acmart}
\AtBeginDocument{%
  }

\setcopyright{acmlicensed}
\copyrightyear{2025}
\acmYear{2025}
\acmDOI{XXXXXXX.XXXXXXX}
\acmConference[RecSys '25]{The 19th ACM Recommender Systems Conference}{September 22--26,
  2025}{Prague, Czech Republic}
\acmBooktitle{RecSys 2025 EARL Workshop,
 September 22--26, 2025, Prague, Czech Republic}
\acmISBN{978-1-4503-XXXX-X/2018/06}

\usepackage{multirow}
\usepackage{subcaption}

\begin{document}

\title{LLM-based Relevance Assessment for Web-Scale Search Evaluation at Pinterest}

\author{Han Wang}
\authornote{Equal contribution}
\affiliation{%
  \institution{Pinterest}
  \city{San Francisco, CA}
  \country{USA}}
\email{hanwang@pinterest.com}

\author{Alex Whitworth}
\authornotemark[1]
\affiliation{%
  \institution{Pinterest}
  \city{San Francisco, CA}
  \country{USA}}
\email{awhitworth@pinterest.com}

\author{Pak Ming Cheung}
\affiliation{%
  \institution{Pinterest}
  \city{San Francisco, CA}
  \country{USA}}
\email{pcheung@pinterest.com}

\author{Zhenjie Zhang}
\affiliation{%
  \institution{Pinterest}
  \city{San Francisco, CA}
  \country{USA}}
\email{zhenjiezhang@pinterest.com}

\author{Krishna Kamath}
\affiliation{%
  \institution{Pinterest}
  \city{San Francisco, CA}
  \country{USA}}
\email{kkamath@pinterest.com}

\renewcommand{\shortauthors}{Wang et al.}

\begin{abstract}
Relevance evaluation plays a crucial role in personalized search systems to ensure that search results align with a user’s queries and intent. While human annotation is the traditional method for relevance evaluation, its high cost and long turnaround time limit its scalability. In this work, we present our approach at Pinterest Search to automate relevance evaluation for online A/B experiments using fine-tuned LLMs. We rigorously validate the alignment between LLM-generated judgments and human annotations, demonstrating that LLMs can provide reliable relevance measurement for experiments while greatly improving the evaluation efficiency. Leveraging LLM-based labeling further unlocks the opportunities to expand the query set, optimize sampling design, and efficiently assess a wider range of search experiences at scale. This approach leads to higher-quality relevance metrics and significantly reduces the Minimum Detectable Effect (MDE) in online experiment measurements.
\end{abstract}



\received{04 September 2025}

\maketitle

\section{Introduction}

Search relevance measures how well search results align with user's search query. For personalized search systems, evaluating relevance is crucial to ensure that displayed contents are pertinent to the user's information needs, rather than over-relying on user's past engagement. 
Online A/B experiments are ubiquitous in industry recommendation systems to measure the impact on user metrics. For search systems, it is essential to track changes in whole-page search relevance during experimentation.

Relevance evaluation typically relies on human annotations. However, this approach suffers from high costs, long turnaround time, and limited scalability, which can ultimately compromise the quality of relevance metrics. In recent years, with the advancement of Large Language Models (LLMs), there is a growing interest in automating relevance assessment in the Information Retrieval (IR) community. There have been several works exploring the feasibility of using LLMs to supplement laborious human labeling efforts \cite{faggioli2023perspectives,thomas2024large,upadhyay2024umbrela}.

At Pinterest, search is one of the key surfaces where users discover inspiring content that meets their information needs. We track whole-page relevance to ensure a high-quality user experience.
In this paper, we present our methodology at Pinterest Search to leverage LLMs for assessing semantic relevance in online A/B experiments. We fine-tune open-source LLMs on relevance prediction tasks using human-annotated labels, then utilize the fine-tuned LLMs to evaluate the ranking results across experimental groups in online A/B experiments. This approach not only significantly reduces labeling costs and improves evaluation efficiency, but also unlocks opportunities to further improve metric quality by scaling up the query sets and refining the sampling design. 
We hope that this work is useful to other industry practitioners looking to improve the efficiency of relevance measurement. 

We highlight our contributions as follows: 
\begin{itemize}
    \item We present the application of LLM-based relevance assessment at Pinterest Search, and outline the practical considerations and validation steps performed prior to adopting LLM-based automated relevance labeling.
    \item We demonstrate that fine-tuning small-sized LLMs on human-annotated relevance labels produces highly effective relevance models, with generated metrics closely aligning with those derived from human judgments. The average error in query-level $sDCG@K$ metric (defined in Section \ref{seq:rel_measure_with_llm}) remains within 0.02.
    \item We highlight that LLM-based relevance assessment allows for expanding the query set and improving the query sampling design, which leads to higher-quality relevance metrics and substantially lowers the Minimum Detectable Effect (MDE) for online experiment evaluation.
\end{itemize}

\section{Related Works}

\subsection{Search Relevance Modeling}

For LLM-based search relevance modeling, there are two types of architectures commonly used in practice: bi-encoders and cross-encoders. Bi-encoder models encode queries and documents separately into a common dense space and then score their relevance using vector dot-product or cosine similarity. These models are typically trained by minimizing the contrastive loss with in-batch negative sampling \cite{karpukhin2020dense,ma2023fine}. While bi-encoder models are efficient for retrieval tasks, they are limited in their ability to capture complex interactions between queries and documents. In contrast, cross-encoder models encode queries and documents jointly, hence can better capture the interaction between queries and documents. Various loss functions are used in the literature for training cross-encoder relevance models, such as pointwise classification loss \cite{nogueira2020document}, pairwise or listwise ranking loss \cite{zhuang2023rankt5}, and contrastive loss \cite{ma2023fine}. In this work, we adopt a cross-encoder architecture and fine-tune the LLMs by minimizing the point-wise multi-class classification loss. The fine-tuned LLMs are then applied to whole-page relevance assessment at Pinterest Search.

\subsection{LLMs for Relevance Judgments}

Recent studies have shown the potential of using LLMs for automating relevance assessments in Information Retrieval (IR), as part of the broader “LLM-as-a-judge” paradigm \cite{gu2024survey,zheng2023judging}. Initial explorations by \citet{faggioli2023perspectives} examined the application of LLMs to relevance judgment, providing insights into human-LLM collaboration. Subsequent work by \citet{thomas2024large} and \citet{upadhyay2024umbrela} further demonstrated the capability of LLMs to accurately predict searchers' preferences and generate human-quality relevance labels. 
Most current research investigates various prompting techniques for LLM-based relevance labeling, including zero-shot \cite{upadhyay2024umbrela,arabzadeh2025benchmarking,upadhyay2024large} and few-shot \cite{macavaney2023one,pires2025expanding} approaches, with a primary focus on evaluating their alignment with human labeling. Later, \citet{meng2025query} and \citet{abbasiantaeb2024can} showed that fine-tuning open-source LLMs on human-labeled relevance judgments is crucial for obtaining more reliable relevance predictions, suggesting that fine-tuning much smaller LLMs can yield more effective relevance prediction than few-shot prompting with significantly larger models.

Unlike most existing work that primarily compares different LLMs and prompting techniques \cite{meng2025query,abbasiantaeb2024can} or explores human-LLM collaboration \cite{shankar2024validates}, our work presents a real-world application of LLM-based relevance assessment to measure relevance change in online experiments at Pinterest Search. We discuss the practical considerations and validation steps, and highlight the opportunities for improving relevance measurement driven by this approach.

\section{Methodology}

\subsection{Problem Statement}

At Pinterest, we measure the semantic relevance between queries and Pins\footnote{Pins on Pinterest are rich multimedia entities that feature images, videos and other content, often linked to external webpages or blogs.} using a 5-point guideline: Highly Relevant (L5), Relevant (L4), Marginally Relevant (L3), Irrelevant (L2), and Highly Irrelevant (L1). 
We use this guideline to measure the whole-page relevance for our search system. 
For online experiments with new search ranking models, assessing relevance across experimental groups is essential for monitoring the relevance impact. 
The relevance measurement procedure consists of the following components:
\begin{itemize}
    \item Data collection: define and collect a representative sample.
    \item Relevance Labeling: rate each element of the sample.
    \item Metrics calculation: aggregate rated elements to get the overall relevance metrics.
\end{itemize}

These measurements have historically been limited by the low availability of human labels and the high marginal cost associated with generating them. This led to measurement designs and sample sizes that could only detect significant topline metric movements, but were insufficient to measure heterogeneous treatment effects or small topline effects. Next, we describe how we leverage LLMs to scale the labeling capabilities and address these bottlenecks.


\subsection{Fine-tuned LLMs as Relevance Model}

\begin{figure}[h]
  \centering
  \includegraphics[width=0.5\linewidth]{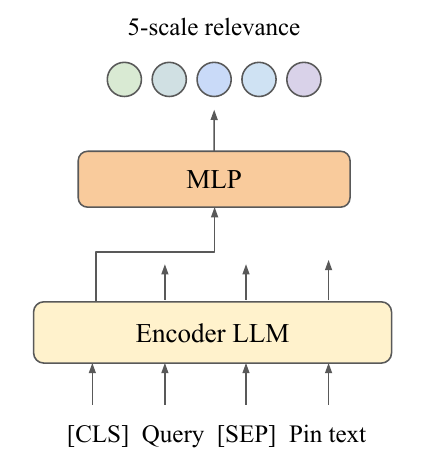}
  \caption{The cross-encoder architecture in the relevance teacher model. Take the encoder language models (e.g., BERT-based models) for illustration.}
  \Description{The figure illustrates the cross-encoder architecture in the relevance teacher model. We take the encoder language models (e.g., BERT-based models) for example.}
  \label{fig:model_architecture}
\end{figure}

A cross-encoder model architecture is utilized to predict a Pin's relevance to a given query, as illustrated in Figure \ref{fig:model_architecture}. We fine-tune open-source LLMs on human-annotated data to optimize their performance on relevance prediction task. The training dataset contains approximately 2.6 million human-annotated query and Pin pairs, and model evaluation is performed on a held-out test split to prevent overfitting. 
To support search queries and Pins across multiple languages, we leverage multilingual LLMs to take advantage of their cross-lingual transfer capabilities. We formalize the relevance prediction as a multiclass classification problem based on the 5-point relevance guideline, minimizing the point-wise cross-entropy loss during training.

\begin{figure*}[h!]
  \centering
  \includegraphics[width=0.9\linewidth]{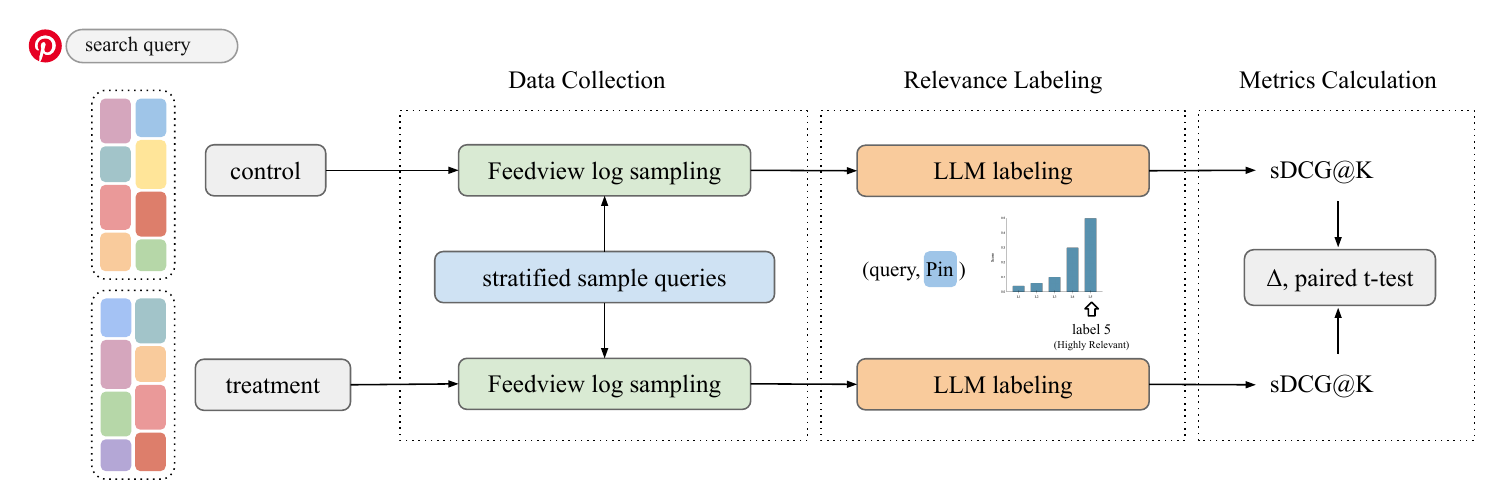}
  \caption{Components of LLM-based relevance measurement at Pinterest Search.}
  \Description{Components of LLM-based relevance measurement at Pinterest Search: data collection, relevance labeling, and metrics calculation.}
  \label{fig:relevance_measurement}
\end{figure*}

To effectively represent each Pin for relevance prediction, we leverage a comprehensive set of textual features, including Pin titles and descriptions, BLIP image captions \cite{li2022blip}, linked page titles and descriptions, user-curated board titles where the Pin has been saved, and highly-engaged queries associated with the Pin. These features together form a robust text representation crucial for accurate relevance assessment. 

We experiment with various pre-trained language models, including multilingual BERT$_{base}$ \cite{devlin2018bert}, T5$_{base}$ \cite{2020t5}, mDeBERTaV3$_{base}$ \cite{he2021debertav3}, XLM-RoBERTa$_{large}$ \cite{conneau2019unsupervised}, and Llama-3-8B \cite{llama3modelcard}. The comparative performance for these models and ablation studies on Pin text features can be found in \citet{wang2024improving}. 
We then use the fine-tuned model to generate 5-dimensional relevance scores and use the label corresponding to the highest score (argmax) for relevance assessment.

\subsection{Stratified Sampling Design}

LLM labeling significantly reduces relevance labeling costs as well as labeling time, which enables much larger sampling designs. Therefore, we propose a stratified query sampling design that enables measurement of heterogeneous treatment effects and reduces minimum detectable effects (MDEs) by an order of magnitude. Prior to LLM labeling, stratified query sampling with human annotations was impractical, as it required a large number of queries to adequately represent each fine-grained stratum.

Stratification plays an important role in sampling-based measurement. First, stratification ensures the sample population is representative of the whole population. In addition, if the strata are chosen such that each stratum is relatively homogeneous, variance reduction can be achieved \cite{miratrix2013adjusting}. From a statistical perspective, 
consider the whole-page relevance score as a random variable $Y$ with mean $\mu$ and variance $\sigma^2$. The variance of the sample mean $\bar Y$ is given by $V(\bar Y) = \sigma^2 / N$, where $N$ is the sample size.
If $K$ strata are chosen such that each stratum has a different mean, $V(\bar{Y})$ can be decomposed into within-strata and between-strata variance (see Equation \ref{eq:strata_variance}), the latter of which can be removed due to stratified sampling \cite{xie2016improving},
\begin{equation}
    V(\bar{Y})=\sum_{k=1}^K \frac{n_k}{N}\sigma_k^2+\sum_{k=1}^K \frac{n_k}{N}(\mu_k-\mu)^2 \geq \sum_{k=1}^K \frac{n_k}{N}\sigma_k^2 = V(\bar{Y}_{\mathrm{strata}}),
    \label{eq:strata_variance}
\end{equation}
where $n_k$, $\mu_k$, $\sigma_k^2$ denote the sample size, mean, and variance of each stratum. Therefore, by using a stratified sampling design, we can achieve reduced variance and, as a result, reduced MDEs.

To determine the query strata, we evaluated multiple strata choices. While BERTopic \cite{grootendorst2022bertopic} produced the largest variance reduction, we adopted the in-house query-to-interest model based on DistilBERT \cite{sanh2019distilbert}, due to ease of integration with existing systems and scalability.

\subsection{Relevance Measurement with LLMs}
\label{seq:rel_measure_with_llm}

To measure the relevance impact of an A/B experiment on search ranking, we take a stratified sample of paired search queries from control and treatment experiment groups, ensuring that the sample is representative of overall user usage \cite{thompson2012sampling}. The use of paired samples blocks between-query differences,
an important source of variation in experiment measurement.

For each query in our paired sample, we retain the top $K$ search results and generate LLM-based relevance labels. We then compute $sDCG@K$ for each query and aggregate query-level metrics to derive topline experiment metrics. The $sDCG@K$ metric is a variant of the standard $nDCG@K$, where we assume an infinite supply of highly relevant (L5) documents (see Equation \ref{eq:sdcg}). We use $K=25$ throughout our evaluation. 
\begin{equation}
\label{eq:sdcg}
    sDCG@K=\frac{\sum_{k=1}^K L_k / log_2(1+k)}{\sum_{k=1}^K 5 / log_2(1+k)}, \text{ }L_k\in\{1,2,3,4,5\}.
\end{equation}

Lastly, we calculate heterogeneous effects by query popularity and query interest (e.g. beauty, women’s fashion, art, etc), utilizing a Benjamini-Hochberg procedure \cite{benjamini1995controlling} to control the false discovery rate. 
The LLM-based relevance measurement procedure at Pinterest Search is illustrated in Figure \ref{fig:relevance_measurement}. 


\section{Results}

\begin{table*}[h!]
  \caption{LLM vs human labels alignment in query-level sDCG@K for different query popularity segments in US market relevance evaluation. }
  \label{tab:us_label_alignment}
  \scalebox{0.95}{\begin{tabular}{l|cc|cccc}
    \toprule
    \multirow{2}{8em}{Query Popularity} & \multirow{2}{6em}{Kendall's $\tau$} & \multirow{2}{6em}{Spearman's $\rho$} & \multicolumn{4}{c}{Query-level Error}   \\
    & & & Mean & P10 & Median & P90 \\
    \midrule
    \midrule
    Overall & 0.652 & 0.817 & 0.005 & -0.047 & 0.002 & 0.059  \\
    \midrule
    Head  & 0.539 & 0.668 & 0.007 & -0.020 & 0.001  & 0.043  \\
    Torso  & 0.600 & 0.764 & 0.008 & -0.036 & 0.004 & 0.057 \\
    Tail  & 0.595 & 0.775 & -0.002 & -0.073 & 0.001 & 0.063 \\
    Single  & 0.620 & 0.801 & 0.004 & -0.068 & 0.003 & 0.075 \\
  \bottomrule
\end{tabular}}
\end{table*}

\begin{table*}[h!]
  \caption{LLM vs human labels alignment in query-level sDCG@K for different query popularity segments in France (FR) and Germany (DE) markets relevance evaluation. }
  \label{tab:fr_de_label_alignment}
  \scalebox{0.95}{\begin{tabular}{ll|cc|cccc}
    \toprule
    \multirow{2}{4em}{Country} & \multirow{2}{8em}{Query Popularity} & \multirow{2}{6em}{Kendall's $\tau$} & \multirow{2}{6em}{Spearman's $\rho$} & \multicolumn{4}{c}{Query-level Error}   \\
    & & & & Mean & P10 & Median & P90 \\
    \midrule
    \midrule
    \multirow{5}{4em}{FR} & Overall & 0.483 & 0.618 & 0.021 & -0.022 & 0.015 & 0.079  \\
    \cmidrule{2-8}
    & Head  & 0.289 & 0.366 & 0.022	& -0.009 & 0.014 & 0.074  \\
    & Torso  & 0.353 & 0.466 & 0.025 & -0.011 & 0.017 & 0.078 \\
    & Tail  & 0.531 & 0.666 & 0.028 & -0.007 & 0.017 & 0.071 \\
    & Single  & 0.572 & 0.743 & 0.015 & -0.057 & 0.014 & 0.086 \\
    \midrule
    \multirow{5}{4em}{DE} & Overall & 0.470 & 0.611 & 0.020 & -0.021 & 0.015 & 0.078 \\
    \cmidrule{2-8}
    & Head  & 0.401 & 0.513 & 0.024 & -0.010 & 0.017 & 0.071  \\
    & Torso  & 0.368 & 0.483 & 0.026 & -0.011 & 0.019 & 0.078 \\
    & Tail  & 0.408 & 0.531 & 0.019 & -0.026 & 0.013 & 0.079 \\
    & Single  & 0.493 & 0.678 & 0.009 & -0.068 & 0.008 & 0.088 \\
  \bottomrule
\end{tabular}}
\end{table*}

We use XLM-RoBERTa$_{large}$ as the LLM backbone for our relevance model. The model is lightweight yet delivers high-quality predictions. Inference runs on a single A10G GPU, allowing us to label 150,000 rows within 30 minutes. While the Llama-3-8B model offers slight improvement in accuracy, its inference time and cost increase by 6 times. Therefore, we pick XLM-RoBERTa$_{large}$ as it offers a good balance between prediction quality and inference efficiency. 
Results are presented to answer the following three research questions (RQs): 
\begin{itemize}
    \item \textbf{RQ1}: Can LLM-based relevance assessment provide reliable metrics for online A/B experiment measurements?
    \item \textbf{RQ2}: Does LLM-based relevance assessment offer greater metric sensitivity compared to human labeling?
    \item \textbf{RQ3}: Is multilingual LLM-based relevance assessment effective for non-English queries as well? 
\end{itemize}

\subsection{RQ1: Alignment with Human Labels}
\label{sec:rq1_us}

We conducted a rigorous validation of the metrics derived from LLM labeling
by comparing with human labels. The human labels were collected by instructing annotators to assess the relevance of each search query and Pin pair. 
The exact match rate between LLM and human labels is 73.7\%, with 91.7\% of ratings differing by no more than one point. These results underscore the high alignment between the relevance labels produced by LLMs and those from human annotators. 

Following the established best practices in literature \cite{meng2025query,abbasiantaeb2024can} to measure alignment between LLMs and human labels, we compute and report the rank-based correlation Kendall’s $\tau$ and Spearman’s $\rho$ to assess the correlation between the two rankings at query-level sDCG@K metric. To understand the performance on queries with different popularity, we also categorize the queries into 4 popularity segments based on search volume \cite{jain2010organizing}: head, torso, tail, and single\footnote{Single queries are defined as those that receive fewer than 10 searches.}.
The results are summarized in Table \ref{tab:us_label_alignment}. For all query popularity segments, we achieve Kendall’s $\tau>0.5$ and Spearman’s $\rho>0.65$, indicating a strong alignment across all segments.

\begin{figure}[h]
  \centering
  \includegraphics[width=\linewidth]{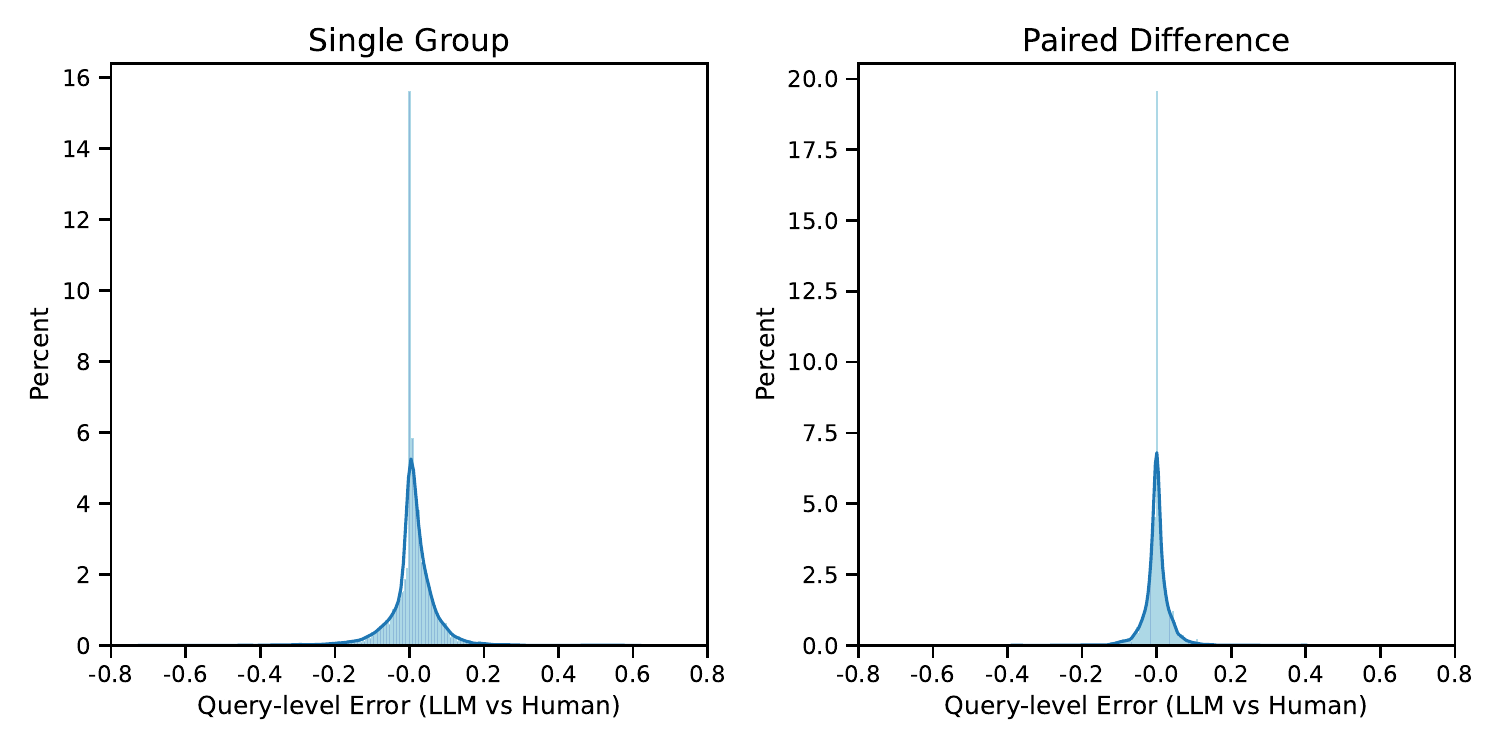}
  \caption{Query-level $\boldsymbol{sDCG@K}$ error distribution for single group (left) and paired differences (right) in US market relevance evaluation. }
  \Description{Query-level $\boldsymbol{sDCG@K}$ error distribution for single group (left) and paired differences (right) in US market relevance evaluation.}
  \label{fig:us_error}
\end{figure}

In addition to Kendall’s $\tau$ and Spearman’s $\rho$, we also validate the query-level sDCG@K error distribution. Here, the error refers to the difference between the sDCG@K metric derived from LLM labels and human labels. According to Table \ref{tab:us_label_alignment}, the overall error is below 0.01, with the 10-th and 90-th percentiles falling within the range of [-0.1, 0.1]. We also visualize the error distribution in Figure \ref{fig:us_error}. The error is tightly centered around 0, indicating its negligible magnitude and that the average bias will approach 0 as the size of the query set grows. 

For experimental evaluation, we need to calculate the metric difference between the control and treatment groups. Therefore, we also validate how well these metric differences align in paired comparison. As shown on the right-hand side of Figure \ref{fig:us_error}, the errors in paired differences are even more centered around 0 with lighter tails, indicating that LLM-based labeling provides highly reliable estimates of paired differences for A/B experiment assessment.\footnote{Note that, despite the low bias, we continue to use human-annotated relevance labels for evaluating online experiments that involve relevance ranking model changes in order to avoid potential bias.}



\subsection{RQ2: Metrics Sensitivity and MDEs}

We evaluate the impact of these changes on experiment sensitivity by evaluating the MDEs
for our experimentation system. 
The MDE is the smallest change in a metric that an experiment can reliably detect given the sample size, statistical power, and significance level chosen for the test. 
Since the typical experiment for most online platforms has a small effect, achieving small MDEs is a critical factor in team velocity and shipping new features to our users.

Before the introduction of LLM labeling, relevance measurement had large MDEs, typically ranging from 1.3\% to 1.5\%. These large MDEs were primarily the result of the constraints on our sampling designs imposed by the high cost and time consumption of human labeling. 
The introduction of LLM labeling enabled us to redesign our sampling approach. We increased our sample sizes, moved from simple random sampling (SRS) to stratified sampling, and now use a stratified sampling estimator \cite{thompson2012sampling}. Optimal allocation \cite{neyman_allocation} is used to allocate sample units to strata, which are defined as the intersection of query interest and popularity segment. These changes enabled us to reduce our MDEs to $\leq 0.25\%$.

To attribute the MDE reduction to each source of change, we follow a standard derivation of MDE \cite{ds_power} as a function of the number of queries in the sample ($n$), metric mean ($\hat \mu$), and variance ($\hat \sigma^2$)  assuming standard values for $\alpha = 0.05$ and $\beta = 0.8$,
\begin{equation}
    \mathrm{MDE} = \mathrm{Lift}\% \mathrm{Detectable}=
    \frac{(z_{1-\alpha/2} + z_{\beta}) \times \sqrt{2\hat{\sigma}^2/ n}}{\hat{\mu}}.
    \label{mde_eqn}
\end{equation}
We can therefore express MDE reduction in terms of reduction in $\hat \sigma$ and increased sample size. We present these results in Table \ref{tab:mde}. The vast majority of reduction comes from the variance reduction due to stratification. 
This is consistent with prior findings at Pinterest that most variance in relevance occurs across queries,
largely due to differences in query interest and popularity.

\begin{table}[h!]
  \caption{Improvement in metric sensitivity (MDE) with proposed sampling design and estimator.}
  \label{tab:mde}
  \scalebox{0.95}{
  \begin{tabular}{ccccc}
    \toprule
    Sample Type & Estimator & $n$ & $\hat \sigma$ & Reduction in $\hat \sigma$ \\
    \midrule
    \midrule
    SRS & SRS & 2000 & 0.184 & - \\
    Stratified & Stratified & 2000 & 0.096 & 52\% \\
    Stratified & SRS & 5000 & 0.094 & 51\% \\
    Stratified & Stratified & 5000 & 0.061 & 67\% \\
  \bottomrule
\end{tabular}}
\end{table}

\subsection{RQ3: Performance on Non-English Queries}

We fine-tuned multilingual LLMs on human-annotated data, with the majority of query-Pin pairs in English. As a result, careful validation is required for non-English queries to extend LLM-based relevance assessment to those queries. For this analysis, we focus on France (FR) and Germany (DE) markets.

\begin{figure}[h!]
  \begin{center}
  \begin{subfigure}{0.95\linewidth}
    \includegraphics[width=\linewidth]{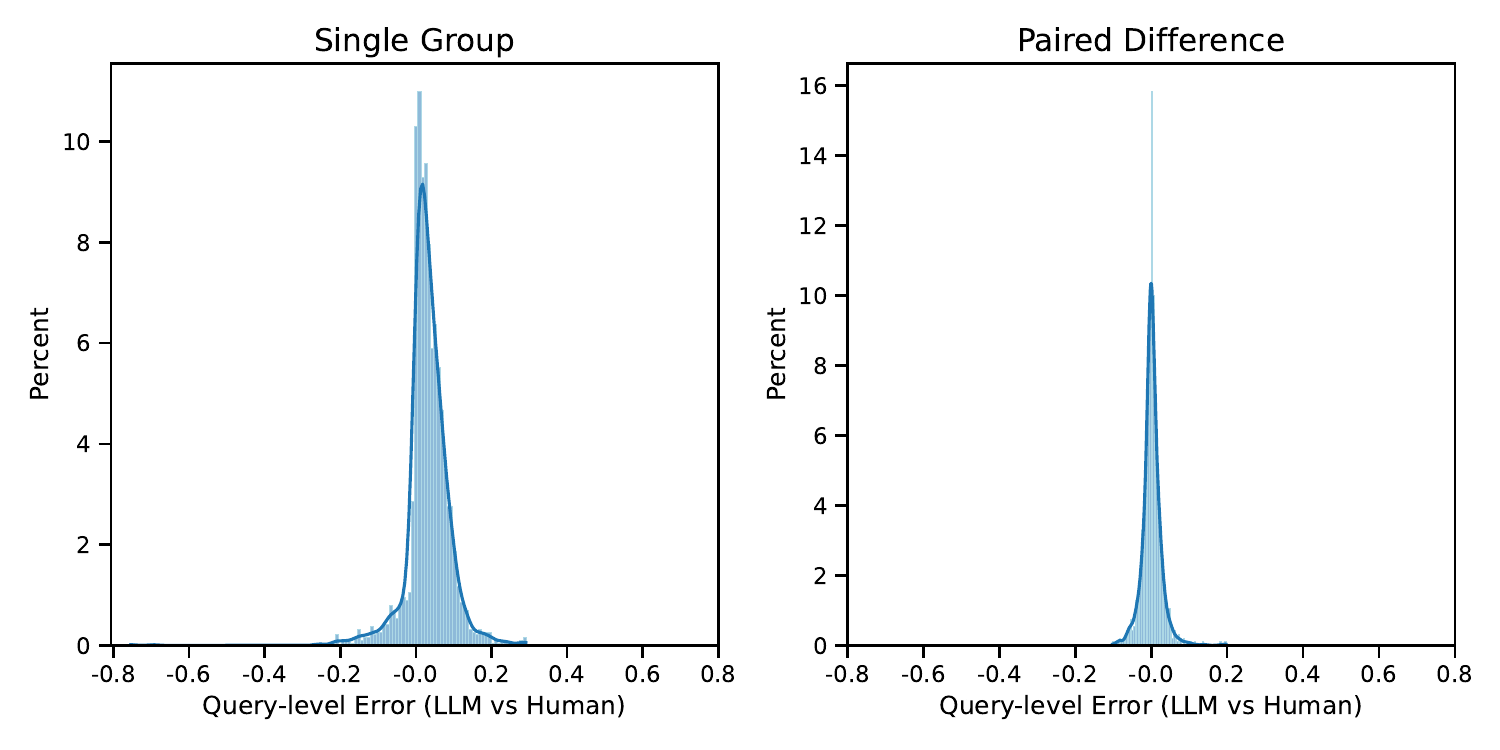}
    \end{subfigure}
  \begin{subfigure}{0.95\linewidth}
    \includegraphics[width=\linewidth]{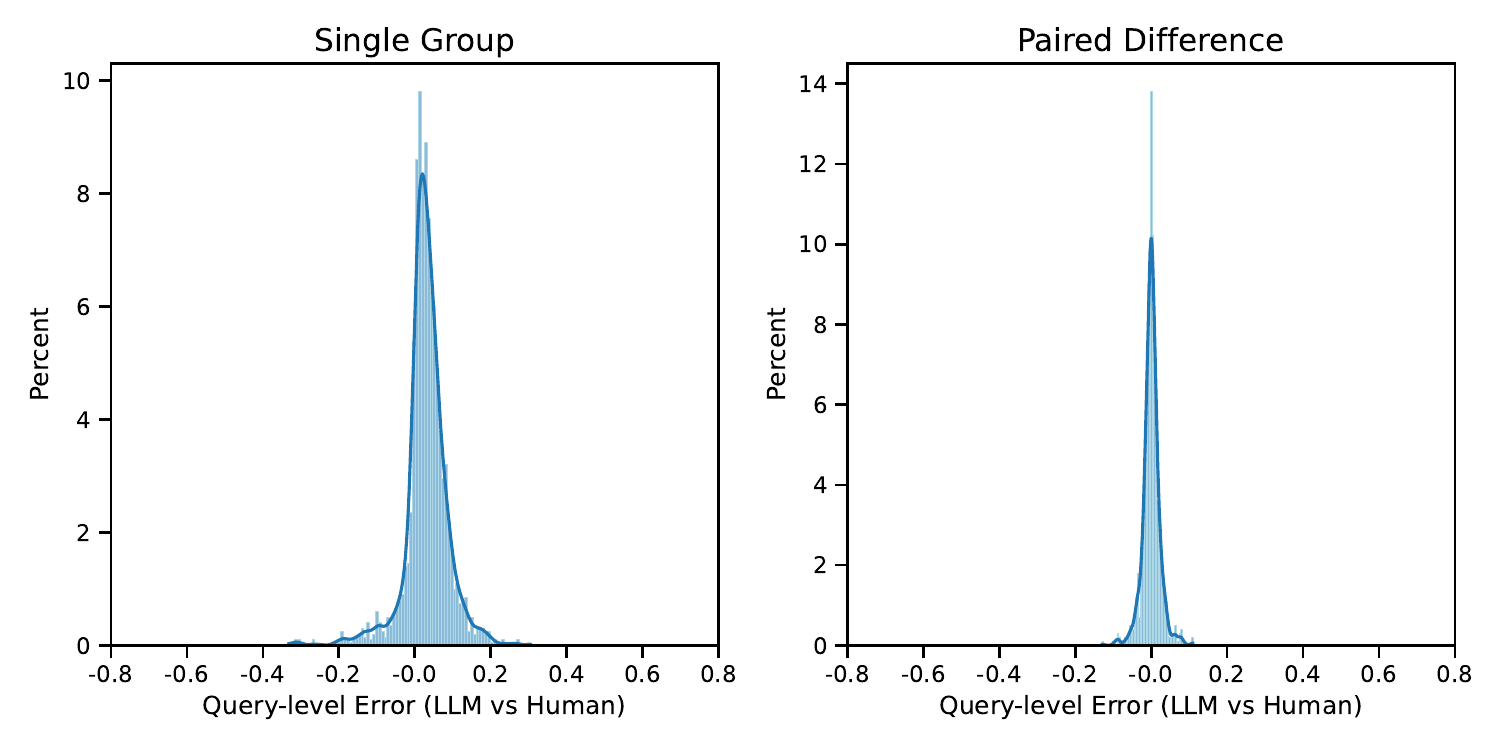}
    \end{subfigure}
\end{center}
  \caption{Query-level $\boldsymbol{sDCG@K}$ error distribution for single group (left) and paired differences (right) in France (top) and Germany (bottom) markets relevance evaluation.}
  \Description{Query-level $\boldsymbol{sDCG@K}$ error distribution for single group (left) and paired differences (right) in France (top) and Germany (bottom) markets relevance evaluation.}
  \label{fig:fr_de_error}
\end{figure}

The query-level metric alignment is summarized in Table \ref{tab:fr_de_label_alignment}. The overall Kendall’s $\tau$ and Spearman’s $\rho$ are approximately 0.47 and 0.61, respectively.  While these rank-based correlations are lower than those observed for English queries, they are still considered strong according to existing literature. The distribution of query-level metric errors is shown in Figure \ref{fig:fr_de_error}. Similar to the results of the US market, the errors are tightly concentrated around 0 for both countries, indicating a low average bias, with an even smaller bias for paired differences. These results provide confidence that the LLM-based relevance assessment is also suitable for non-English queries. Expanding relevance evaluation to countries beyond the US leads to further reductions in labeling costs and improvements in evaluation efficiency.


\section{Conclusion and Future Work}

In this work, we explore the use of LLM-based relevance labeling to generate query-level relevance metrics for online A/B experiments evaluation. We demonstrate that fine-tuned LLMs achieve low bias on query-level $sDCG@K$ metrics and paired differences. Transition to LLM-based relevance assessment enables us to scale up the evaluation query set and redesign the sampling strategy to improve the quality of relevance metrics for online experiment evaluation. We have successfully deployed the LLM-based relevance assessment at Pinterest Search, significantly reducing the manual annotation costs and turnaround time, while achieving an order of magnitude reduction in MDEs for improved detection of relevance shifts. Future work will explore using Visual Language Models (VLMs) to better leverage raw images for relevance prediction. Additionally, the observed performance gap with non-English queries highlights opportunities to further improve the multilingual capabilities of our LLM-based relevance model. We leave it for future work.


\begin{acks}
We would like to thank Kurchi Subhra Hazra for the support throughout this project;  Mukuntha Narayanan and Jinfeng Rao for collaboration on developing the LLM-based relevance model; and Maggie Yang, Maria Alejandra Morales Gutierrez, Miguel Madera, Pedro Sanchez, Jorge Amigon, Francisco Navarrete for their assistance with LLM-labeling integration.
\end{acks}



\bibliographystyle{ACM-Reference-Format}
\bibliography{main-reference}

\appendix

\end{document}